\documentclass[10pt,a4paper]{article}
\usepackage[utf8]{inputenc}
\usepackage[english]{babel}
\usepackage[T1]{fontenc}
\usepackage{amsmath}
\usepackage{amsfonts}
\usepackage{amssymb}
\usepackage{makeidx}
\usepackage{graphicx}
\usepackage{url}
\usepackage{siunitx}
\usepackage{epsfig}
\usepackage{multirow}
\usepackage{helvet}
\usepackage[table]{xcolor}
\usepackage[version=4]{mhchem}
\usepackage[inner=1.5cm,outer=1.5cm,top=1.5cm,bottom=2cm]{geometry}
\usepackage[font={small}]{caption}
\usepackage[square,numbers]{natbib}
\usepackage{wrapfig}
\usepackage{natbib}
\usepackage{titlesec}
\usepackage{multicol}
\usepackage{xspace}
\usepackage{sectsty}
\usepackage{chngcntr}
\usepackage{hyperref}

\sectionfont{\fontsize{11}{10}\selectfont}
\subsectionfont{\fontsize{10}{9}\selectfont}

\usepackage[blocks]{authblk}

\usepackage{bbding} 
\definecolor{blue}{HTML}{1f77b4}
\definecolor{orange}{HTML}{ff7f0e}
\definecolor{green}{HTML}{2ca02c}
\definecolor{ashgrey}{rgb}{0.7, 0.75, 0.71}

\newcommand{\sco}{$\ce{SrCrO_3}$\xspace}

\newcommand{\rco}[1]{$\ce{#1CrO_3}$\xspace}

\newcommand{\nao}{$\ce{NdAlO_3}$\xspace}
\newcommand{\naon}{$\ce{NdAlO_3}$(001)\xspace}
\newcommand{\lao}{$\ce{LaAlO_3}$\xspace}
\newcommand{\laon}{$\ce{LaAlO_3}$(001)\xspace}

\newcommand{\ngo}{$\ce{NdGaO_3}$\xspace}
\newcommand{\ngon}{$\ce{NdGaO_3}$(110)\xspace}
\newcommand{\sto}{$\ce{SrTiO_3}$\xspace}
\newcommand{\ston}{$\ce{SrTiO_3}$(001)\xspace}
\newcommand{\dso}{$\ce{DyScO_3}$\xspace}
\newcommand{\dson}{$\ce{DyScO_3}$(110)\xspace}

\newcommand{\lsatn}{$\ce{(LaAlO_3)_{0.3}(Sr_2TaAlO_6)_{0.7}}$(001)\xspace}

\newcommand{\msr}{$\mu$SR\xspace}

\def\snao{nao}\def\slao{lao}\def\snsat{nsat}\def\sngo{ngo}\def\slsat{lsat}\def\ssto{sto}\def\sdso{dso}
\newcommand{\strain}[1]{\def\x{#1}\ifx\x\snao{-1.82\%}\else
\ifx\x\slao{-0.83\%}\else
\ifx\x\snsat{0.63\%}\else
\ifx\x\sngo{1.11\%}\else
\ifx\x\slsat{1.31\%}\else
\ifx\x\ssto{2.25\%}\else
\ifx\x\sdso{3.18\%}\else
{\textbf{?strain?}}
\fi\fi\fi\fi\fi\fi\fi}

\makeatletter
\renewcommand{\@maketitle}{
	\newpage
	\null
	\vskip 2em%
	\begin{center}%
		{\fontsize{16pt}{16pt}\selectfont \@title \par}
		\vspace{0.5cm}
		{\small \@author \par}
    \vspace{0.4cm}
     (Last update : \@date)
	\end{center}%
	\par} \makeatother
\title{\vspace{-2cm}\textbf{Strain-Induced Metal-to-Insulator Transition in Antiferromagnetic SrCrO$_3$ Thin Films}}
\author[1]{\textbf{Simon Jöhr}}
\author[2,6]{Alberto Carta}
\author[3]{Juan Moreno}
\author[5]{Andreas Suter}
\author[5]{Zaher Salman}
\author[2]{Anwesha Panda}
\author[1]{Jonathan Spring}
\author[7]{Gabriele De Luca}
\author[8]{Javier Herrero-Martin}
\author[9]{Bernat Mundet}
\author[4]{Cinthia Piamonteze}
\author[10]{Marc Gabay}
\author[2]{Claude Ederer}
\author[3]{Marta Gibert}

\affil[1]{\textit{University of Zurich, Zurich, Switzerland}}
\affil[2]{\textit{Materials Theory, ETH Zurich, Switzerland}}
\affil[3]{\textit{Technical University of Vienna, Vienna, Austria}}
\affil[4]{\textit{PSI Center for Photon Science, 5232 Villigen, Switzerland}}
\affil[5]{\textit{PSI Center for Neutron and Muon Sciences, 5232 Villigen PSI, Switzerland}}
\affil[6]{\textit{Paul Scherrer Institut, 5232 Villigen, Switzerland}}
\affil[7]{\textit{Institut de Ciència de Materials de Barcelona (ICMAB-CSIC), 08193 Cerdanyola del Valles, Spain}}
\affil[8]{\textit{ALBA Synchrotron Light Source, 08193 Cerdanyola del Vallès, Spain}}
\affil[9]{\textit{Institut de Nanosciència i Nanotecnologia (ICN2), 08193 Cerdanyola del Valles, Spain}}
\affil[10]{\textit{University Paris Saclay, Paris, France}}
\date{\today}

\begin{document}
\pagenumbering{arabic}

\maketitle

\begin{abstract}
    Antiferromagnetic (AF) metals are rare, yet they combine properties attractive for spintronic devices like robustness against stray fields and electrical readout. Among  AF metal oxide candidates, SrCrO$_3$ remains largely unexplored due to its notoriously difficult synthesis. In this paper, we demonstrate the growth of high-quality SrCrO$_3$ thin films by magnetron sputtering on substrates that impose a wide range of tensile and compressive strains. Muon spin relaxation experiments, supported by x-ray magnetic dichroism, unveil the emergence of an AF phase with dilute magnetic disorder at low temperatures, while resistivity measurements confirm the simultaneous metallic ground state of SrCrO$_3$ under low strain. As both compressive and tensile strain increase, a metal-to-insulator transition is induced in the films, while the onset of the magnetic transition temperature remains unchanged. Moreover, an intriguing resistivity upturn, accompanied by a change in the dominant charge-carrier type, occurs at a temperature that correlates with strain. These observations suggest a complex strain-dependent band structure, with strain-induced Jahn-Teller distortions or tilting of the CrO$_6$ octahedra that emerge depending on the sign of the strain, as inferred from density functional theory calculations. 
\end{abstract}

\begin{multicols}{2}


\section{Introduction}
Antiferromagnetic (AF) materials are strong contenders for next-generation spintronic applications due to their unique combination of robustness to external magnetic fields, lack of stray fields and ultrafast dynamics \cite{gomonay_antiferromagnetic_2018, siddiqui_metallic_2020, arpaci_observation_2021}. 
Recent advances in antiferromagnetic spintronics have enabled sophisticated control mechanisms through electrical, optical, and strain-based approaches \cite{bai_antiferromagnetism_2022, macdonald_antiferromagnetic_2011, jungwirth_antiferromagnetic_2016}.
However, while effects such as spin transfer torque, giant magnetoresistance, and spin valve operation are well established in ferromagnetic systems \cite{mishra_emerging_2021}, writing and reading a magnetic state in AF materials is still very challenging. These tasks are typically performed through measurements of the anisotropic magnetoresistance signal, which typically relies on metallic transport properties \cite{shim_spin-polarized_2024}. Therefore, metallic antiferromagnets are gaining increasing attention as they offer a unique platform where charge, spin, optical effects, and magnetization dynamics can strongly interact.\\ 
\\
Within transition metal oxides, the combination of antiferromagnetism and metallicity is particularly elusive, as AF order is typically associated with insulating behaviour. $\ce{RuO_2}$ is an interesting case that, besides being an itinerant antiferromagnet, has been proposed to be an altermagnetic system \cite{jungwirth_symmetry_2026}.
Perovskite nickelates, which undergo a  metal-insulator and AF transition as a function of temperature, can also be tuned into an AF metal through doping \cite{song_antiferromagnetic_2023}.\\
\\
The relatively unexplored alkaline earth perovskite chromates with the formula \rco{A} stand out as other potential AF metal candidates, particularly for A = Ca and Sr. While the reported physical properties of \rco{Ca} clearly point to an AF metal \cite{komarek_cacro3_2008, goodenough_band_1968,weiher_magnetic_1971}, the case of \sco is more ambiguous as its physical properties remain debated in contradictory reports. Since its first synthesis in 1967, bulk \sco has been reported as metallic \cite{chamberland_preparation_1967, williams_charge_2006} or insulating \cite{zhou_anomalous_2006},  paramagnetic \cite{chamberland_preparation_1967, zhou_anomalous_2006, ortega-san-martin_microstrain_2007}, ferromagnetic \cite{williams_charge_2006} or exhibiting AF order below a Néel temperature (T$_\text{Néel}$) of 35-40 K only in the low-temperature tetragonal phase\cite{ortega-san-martin_microstrain_2007, komarek_magnetic_2011}, while cubic \cite{chamberland_preparation_1967, ortega-san-martin_microstrain_2007}, tetragonal \cite{ortega-san-martin_microstrain_2007}, or orthorhombic \cite{ding_bond_2020} crystal structures have all been reported.
This variability is often attributed to limited crystalline quality and grain boundary effects, which  themselves are due to the challenging synthesis conditions required to stabilize the octahedral coordination for the small $\ce{Cr^{4+}}$-ion~\cite{chamberland_preparation_1967, ortega-san-martin_microstrain_2007, castillo-martinez_revisiting_2007}.\\
\\
Despite the controversies surrounding the experimental characterization of the ground state properties of \sco, first-principle calculations employing density functional theory (DFT) generally agree on an antiferromagnetically ordered  C-AF metallic ground state (magnetic wavevector: $\big[ \tfrac{1}{2}, \tfrac{1}{2}, 0 \big]$) ~\cite{Lee2009_DFTU_OO, Qian2011_weak_correlations, zhang_electronic_2015}, which, in a cubic lattice, corresponds of chains of ferromagnetically aligned spins with an AF alignment between neighboring chains. This magnetic order couples to a tetragonal distortion of the unit cell, causing a contraction of the lattice constant in the z-direction and an elongation in the xy plane~\cite{Lee2009_DFTU_OO, Qian2011_weak_correlations, zhang_electronic_2015}, in agreement with what was reported in Ref.~\cite{ortega-san-martin_microstrain_2007}.\\
\\
The ability to grow epitaxial \sco thin films has generally resulted in higher crystalline quality samples, with the metallic nature clearly shown when \sco is grown under low strain \cite{zhang_reversible_2014, zhang_electronic_2015, zhang_hole-induced_2015, doyle_effects_2024}. A strain-induced transition to an insulating phase is suggested for films grown under higher tensile strain \cite{bertino_strain_2021}.
This behavior is supported by theoretical work, which has shown that \sco becomes Jahn-Teller active under tensile epitaxial strain~\cite{carta_evidence_2022}. The Jahn-Teller distortion induces orbital order at the level of the Cr $d_{xz}$ and $d_{yz}$ orbitals and opens a gap which increases with tensile strain~\cite{carta_evidence_2022}. While the theoretical explanation assumes the presence of magnetic order, in terms of magnetism, \sco films have barely been experimentally explored~\cite{zhang_electronic_2015}.\\
\\
Using a combination of experimental measurements and first-principles-based calculations, we provide new insights into the debated magnetic and transport properties of \sco thin films. Despite the challenging growth of \sco, we successfully synthesized a series of films with strains ranging from $\epsilon=\strain{nao}$ compressive to $\strain{dso}$ tensile strain.
Regarding the transport properties, we demonstrate that low-strain \sco films are metallic and that the resistivity can be strongly tuned via strain-engineering. Specifically, a transition to insulating behaviour is observed at both higher tensile and compressive strain.
Muon spin relaxation and x-ray magnetic dichroic spectroscopy measurements allow us to conclude that the \sco thin films have an AF order with dilute magnetic disorder leading to a percolative formation of spin-ordered regions.  
Nonetheless, together with the transport properties mentioned above, we confirm the AF metal ground state of \sco thin films.
Overall, our work provides a promising route for the exploration of chromate oxide materials and AF metals in general, and motivates further investigations of the spin and orbital dynamics in these materials. 

\section{Results}
\subsection*{Structural Quality}
\begin{figure*}
    \centering
    \includegraphics[width = 0.8\textwidth]{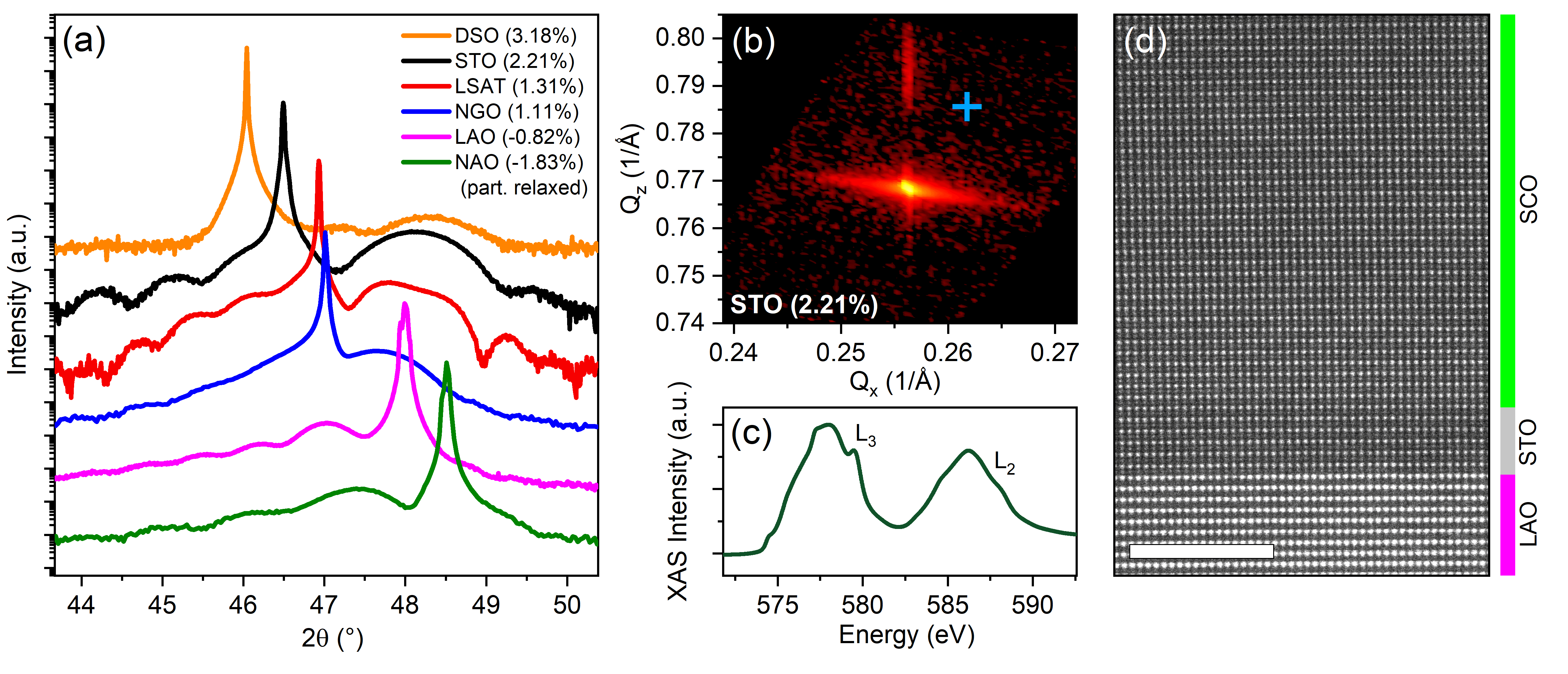}
    \caption{Structural characterization of the \sco thin films. \textbf{(a)} X-ray diffraction (XRD) scans around the (002) pseudo-cubic Bragg peak of $\SI{15}{\nano\meter}$-thick films strained on different substrates. The legend indicates the substrate and the corresponding nominal strain. The finite-size Laue oscillations that are visible in all samples are evidence of their good crystalline quality. Only films grown on \nao appear to be partially relaxed. 
    \textbf{(b)} Reciprocal space map (RSM) performed around the (103) Bragg peak of the \sco film grown on \sto. The blue cross indicates the expected bulk lattice parameter of \sco.
    \textbf{(c)} X-ray absorption spectroscopy (XAS) measurements at the Cr-L$_{3,2}$ edge of a \sco film grown on \lao, revealing a $\ce{Cr^{4+}}$ oxidation state.\cite{zhang_hole-induced_2015, sarma_investigation_1996}
    \textbf{(d)} Scanning transmission electron microscopy (STEM) image of \sco grown on \lao with a 5 to 6 unit cell-thick \sto buffer layer, viewed along the [110] zone axis. The length of the white bar corresponds to $\SI{5}{\nano\meter}$}.
    \label{fig:FIG1}
\end{figure*}

Figure \ref{fig:FIG1}(a) shows the x-ray diffraction patterns of a set of $\SI{15}{\nano\meter}$-thick \sco films grown on substrates imposing either compressive [\naon (NAO), $\epsilon=\strain{nao}$ and \laon (LAO), $\strain{lao}$] or tensile strain [\ngon (NGO), $\strain{ngo}$; \lsatn (LSAT), $\strain{lsat}$; \ston (STO), $\strain{sto}$ and \dson (DSO), $\strain{dso}$]. The strain values are calculated using $\epsilon = a_{sub}/a_{bulk} -1$, where $a_{bulk} = \SI{3.82}{\angstrom}$ is the bulk lattice parameter of \sco \cite{chamberland_preparation_1967} and $a_{sub}$ the corresponding pseudocubic substrate lattice parameter. All films were grown on 5-6 unit-cell-thick SrTiO$_3$ buffer layers, with the exception of those deposited directly on SrTiO$_3$ substrates, and annealed after deposition (see Methods). The SrTiO$_3$ layer was found to improve the topography of the chromate films. Finite-size Laue fringes are visible for all samples, providing evidence of their high crystalline quality. Figure \ref{fig:FIG1}(b) displays a representative reciprocal space map (RSM) acquired around the $(103)$ Bragg reflection of the \sto substrate. 
Additional RSM measurements are shown in Figure \ref{figS:RSM} of the supporting information (SI). All \sco films are coherently strained to the substrate, with the only exception being those grown on \nao. 
A closer look into the structure is provided by the cross-sectional scanning transmission electron microscopy (STEM) image of a \sco film on LaAlO$_3$ substrate shown in Figure \ref{fig:FIG1}(d). This confirms the epitaxial growth and the high crystalline quality of the samples. 
Regarding the electronic structure, x-ray absorption spectroscopy (XAS) measurements at the Cr-L$_{3,2}$ edge on film samples grown on \lao, \ngo and \sto (see Fig. \ref{fig:FIG1}(c) and Fig. \ref{figS:XAS} in the SI) confirm the expected $\ce{Cr^{4+}}$ valence state when compared to previously reported energy scans \cite{sarma_investigation_1996, zhang_hole-induced_2015, doyle_effects_2024}.
Finally, let's point out that although all strained films show high crystallinity and desired oxidation state, surface morphology can differ across samples. Higher tensile strain promotes smooth step-like growth, whereas roughness increases for samples with $\epsilon<\strain{ngo}$ (see Figure \ref{figS:AFM}). This can be attributed to the larger lattice parameter of the oxygen-deficient phase, which is stabilized during growth, prior to annealing.


\subsection*{Transport}
\begin{figure*}[tb]
 \centering
 \includegraphics[width=0.75\linewidth]{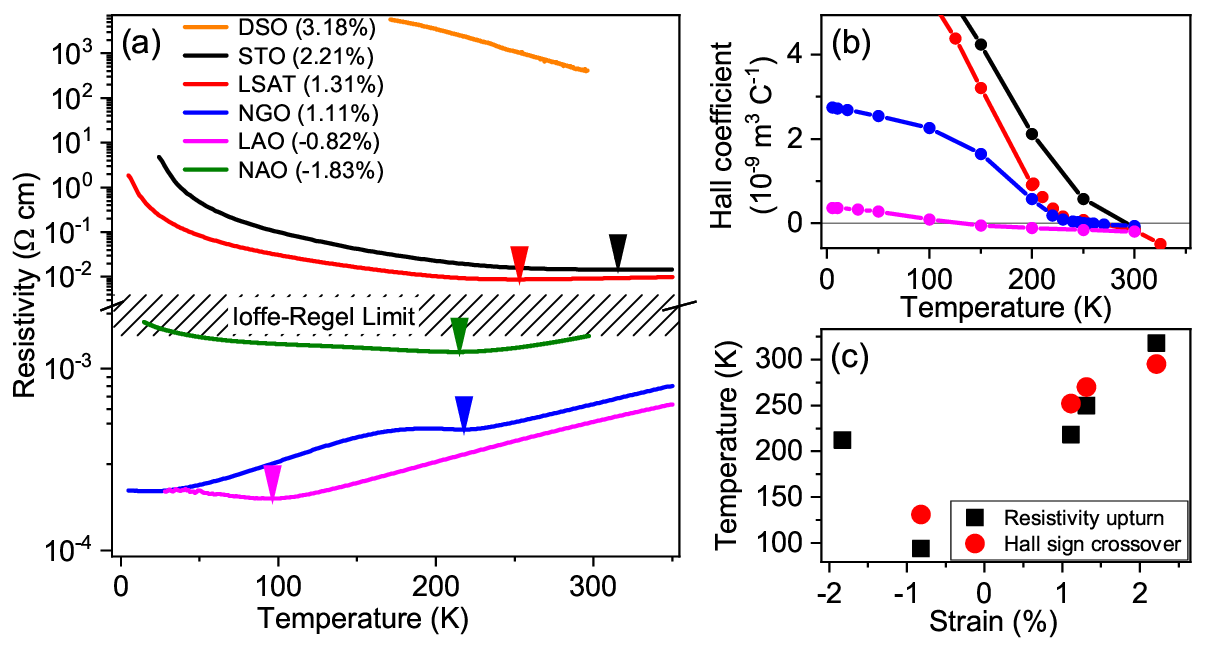}
 \captionof{figure}{Transport properties of \sco thin films as a function of strain.
 \textbf{(a)} Temperature-dependent resistivity of $\SI{15}{\nano\meter}$-thick \sco films. The substrates and corresponding nominal strain values are indicated in the legend.
 The arrows show the resistivity upturn temperatures $T_{upturn}$. The hatched bar indicates the calculated Ioffe-Regel limit as discussed in the main text. 
 \textbf{(b)} Hall coefficients as a function of temperature. 
 \textbf{(c)} $T_\text{upturn}$ (black squares) and $T_\text{cross}$ (red dots) as a function of the strain. 
 }
 \label{fig:FIG2}
\end{figure*}

Figure \ref{fig:FIG2}(a) presents representative resistivity-temperature curves measured for 15 nm-thick \sco thin films. Starting from the lowest absolute strain value, \sco grown on \lao ($\epsilon=-0.82\%$, pink curve) shows a metallic behaviour characterized by a positive slope $\frac{\partial\rho}{\partial T}>0$ down to $T_{upturn}=\SI{100}{\kelvin}$ where a small upturn in resistivity occurs (marked by an arrow). Note that $T_{upturn}$ is defined as the local minima with $\frac{\partial \rho}{\partial T} = 0$.
This sample shows a room temperature resistivity of $\rho(\SI{300}{\kelvin}) = \SI{5.10e-4}{\ohm\centi\meter}$, a value that drops to $\SI{2.11e-4}{\ohm\centi\meter}$ at $\SI{25}{\kelvin}$. This positive resistivity slope at high temperature agrees with previous measurements \cite{zhang_electronic_2015, zhang_hole-induced_2015} and is expected for low strain values, where the \sco films are close to the bulk ground state, often considered metallic \cite{chamberland_preparation_1967, komarek_magnetic_2011, williams_charge_2006}.
A similar behaviour is also observed when \sco is grown on \ngo ($\epsilon = 1.11\%$, blue curve) \cite{doyle_effects_2024}. 
In this case, a first high-temperature resistivity upturn occurs at  $T_\text{upturn} = \SI{218}{\kelvin}$. A metallic behaviour with a positive slope is recovered from $\SI{150}{\kelvin}$ down to a second resistivity upturn located at about $\SI{20}{\kelvin}$. 
The resistivity values of the \sco-on-\ngo film drops by a factor $3$ from $\rho(\SI{300}{\kelvin}) = \SI{6.39e-4}{\ohm\centi\meter}$ to $\rho(\SI{20}{\kelvin}) = \SI{2.12e-4}{\ohm\centi\meter}$.\\
\\
Further increasing the tensile strain leads to a change in the resistivity behaviour. For films grown on LSAT and \sto ($\epsilon = \strain{lsat}$ (red curve) and $\strain{sto}$ (black curve), respectively), a high-temperature phase with positive slope $\frac{\partial\rho}{\partial T}>0$ is only observed down to $T_{upturn}=\SI{250}{\kelvin}$ and $\SI{320}{\kelvin}$, respectively.
Below these upturn temperatures, the \sco thin films show an insulating behaviour, with $\frac{\partial\rho}{\partial T}<0$ and the resistivity increasing by multiple orders of magnitude from $\rho(\SI{300}{\kelvin})=\SI{9.2e-3}{\ohm\centi\meter}$ and $\SI{14.6e-3}{\ohm\centi\meter}$ to $\rho(\SI{20}{\kelvin}) = \SI{0.315}{\ohm\centi\meter}$ and $\SI{5.08}{\ohm\centi\meter}$ for LSAT and \sto, respectively. 
At even higher tensile strain values, the \sco films grown on \dso ($\epsilon=\strain{dso}$, orange line) are characterized by a strongly insulating behaviour with $\rho(\SI{300}{\kelvin})=\SI{404}{\ohm\centi\meter}$. 
No resistivity upturn could be observed in this sample in the measured range (up to $\SI{300}{\kelvin}$).
Turning our attention to the compressive side of the strain spectrum, the \sco films grown on \nao ($\epsilon=\strain{nao}$, green curve) also exhibit a negative slope with respect to temperature at low temperatures, similar to the LSAT and \sto samples, with a resistivity upturn at $T_\text{upturn} = \SI{212}{\kelvin}$ and $\rho(\SI{300}{\kelvin})=\SI{1.5e-3}{\ohm\centi\meter}$. 
Thus, our measurements suggest that \sco also tends toward insulating behaviour with increasing compressive strain.
\\
\\
Collectively, these transport measurements provide compelling evidence for two main points. Firstly, they prove that the ground state of low-strained \sco is that of a metal. Second, they evidence that strain engineering induces a metal-to-insulator transition in \sco films.
We use the Ioffe-Regel criterion to further support the presence of the strain-induced metal-to-insulator transition. Assuming our films are quasi-3D systems, this criterion states that a system localizes strongly if $lk_F < 1$, where $l$ is the mean free path and $k_F$ is the Fermi momentum \cite{ioffe_non-crystalline_1960, calandra_electrical_2002}. 
According to this limit, samples with resistivity higher than the critical resistivity $\rho_c$ are considered to have localized charge carriers,  while those with resistivity below $\rho_c$ exhibit itinerant behaviour. We determine $\rho_c$ using the Drude model and the carrier densities extracted from Hall effect measurements at $\SI{5}{\kelvin}$ for samples grown on \lao and \ngo, and at $\SI{100}{\kelvin}$ for the ones on LSAT and \sto (see Fig. \ref{fig:FIG2}(b) and Fig. \ref{figS:Hall}). 
The calculated $\rho_c$ is indicated as a hatched bar in Figure \ref{fig:FIG2}(a) (see calculations in the SI). Thus, according to the Ioffe-Regel criterion, the metal-to-insulator transition in \sco films occurs between a strain level of $1.1\%$ and $1.3\%$. For the \sco film grown on \nao, the resistivity is slightly below the Ioffe-Regle limit, indicating that \sco is at the edge of localization. However, because this film might be partially relaxed, the actual strain is likely lower than the nominal value and the measured resistivity may not accurately reflect the behaviour of the \sco films at this strain. While the tensile-strain-induced 
transition 
is consistent with previous studies on \sco films \cite{bertino_strain_2021} and recent DFT+U calculations \cite{carta_evidence_2022}, in the next section, we further explore the compressive regime theoretically and we demonstrate that an insulating state can indeed be expected.\\
\\
Before that, it is necessary to address the resistivity upturn occurring in all samples at the temperatures $T_\text{upturn}$, signaled with arrows in Fig. \ref{fig:FIG2}(a) and summarized in Fig. \ref{fig:FIG2}(c) (black squares). We find that $T_\text{upturn}$ correlates well with strain and, moreover, coincides closely with the crossover in the sign of the Hall coefficient at temperatures $T_\text{cross},$ as shown in Figure \ref{fig:FIG2}(c) (red dots). 
At $T_\text{cross}$, the charge carrier majority evolves from electron-like at high temperature to hole-like at low temperature, as evidenced by the temperature-dependent Hall coefficients presented in Figure \ref{fig:FIG2}(b). The field-dependent Hall measurements are shown in the SI Figure \ref{figS:Hall}. 
While the  origin of these Hall sign crossovers remains to be established, a possible explanation could be a structural transition associated with a Fermi surface reconstruction.
This scenario is compatible with the formation of orbital order, made possible by the reorganization of the electronic band structures induced by the strain-dependent JT distortion, as explored in \cite{carta_evidence_2022}, thereby changing the balance of electron and hole pockets in the Fermi surface. However, this is not yet conclusive, and further investigations are needed.

\subsection*{Computational results}
As already mentioned above, 
the insulating phase occurring in \sco thin films under tensile strain has previously been attributed to the emergence of orbital order in combination with a Jahn-Teller (JT) distortion~\cite{carta_evidence_2022}. 
While the material in the bulk at room temperature is reported to be cubic, both the C-AF magnetic state and the tensile epitaxial strain lower the symmetry of the system to tetragonal ($P4/mmm$ space group). This splits the $t_{2g}$ levels of the Cr$^{4+}$ ion, which has a nominal occupation of $d^2$, into a lower lying fully occupied $d_{xy}$ state and two higher lying degenerate $d_{xz}$ and $d_{yz}$ states which share one electron.
As strain increases, a JT distortion becomes favorable, lifting the degeneracy between $d_{xz}$ and $d_{yz}$, resulting in orbital order, \textit{viz.} $d_{xy}^1(d_{xz}d_{yz})^1 \rightarrow d_{xy}^1  d_{xz}^1 d_{yz}^0 \oplus d_{xy}^1 d_{xz}^0 d_{yz}^1$, and a subsequent gap opening. For further details we refer to  Ref.~\cite{carta_evidence_2022}.\\
\\
\begin{figure*}[htbp]
    \centering
    \includegraphics[width=\textwidth]{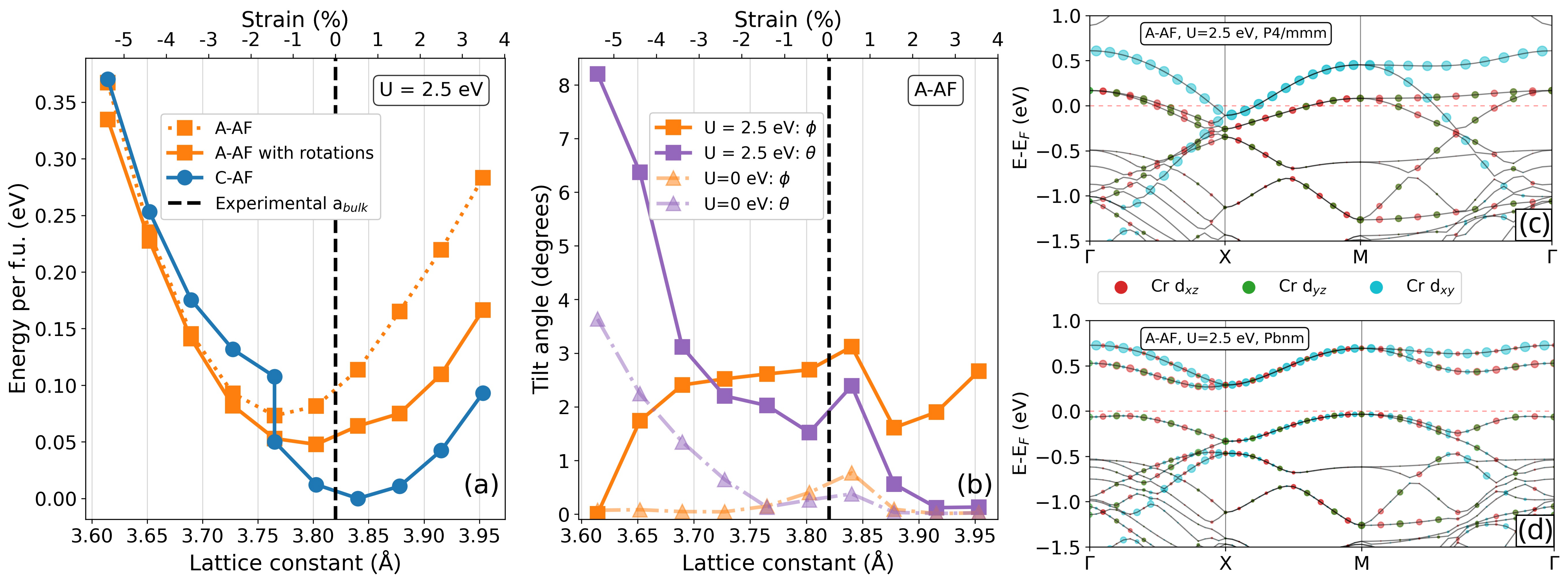}
    \caption{Magnetic phase stability and electronic structure of \sco as a function of strain. (a) Energy per formula unit of several magnetic phases as a function of the in-plane lattice constant 
    for $U$=2.5 eV. Zero energy is set to the minimum of the JT distorted C-AF curve.
    (b) Octahedral tilt angles, $\theta$ and $\phi$, for the A-AF state for both $U=0.0$\,eV and $U=2.5$\,eV.
    The vertical dashed line in (a) and (b) corresponds to the experimental lattice parameter ($a=3.82$ \AA)~\cite{komarek_magnetic_2011}. Epitaxial strain (upper x-axis) is computed with respect to this value.
    (c-d) Band structure including orbital projections on the Cr local majority-spin $t_{2g}$ orbitals at $-1.67$\,\% strain for (c) the $P4/mmm$ tetragonal structure without octahedral tilts and (d) the $Pbnm$ structure with octahedral tilts. The energy $E$ is taken relative to the Fermi level $E_F$.
    }
    \label{fig:THEORY}
\end{figure*}
We now show how insulating behavior may also emerge in the compressive strain regime.
In Fig.~\ref{fig:THEORY}(a), we plot the energy per formula unit against the in-plane lattice constant $a$ for both A-AF (magnetic wavevector [$\frac{1}{2}$,0,0], orange dotted line) and C-AF (solid blue line), calculated within DFT+$U$ with $U=2.5$ eV (see the corresponding methods section for more details).
The reported strain values are computed with respect to the experimental lattice constant $a_\text{bulk}= 3.82$ \AA, which is very close to the pseudocubic lattice constant obtained from the cell volume at the energy minimum of the C-AF phase.
We observe that for higher in-plane lattice constant (tensile regime), C-AF is favored, as was previously reported in~\cite{carta_evidence_2022}. The jump in energy at $a=3.76$ \AA \, corresponds to the emergence of the JT distortion in the C-AF structure for large $a$.
For lower in-plane constants instead (compressive regime), A-AF becomes lower in energy.\\
\\
Since compressive strain lowers the Cr $d_{xz}$ and $ d_{yz}$ orbitals relative to $d_{xy}$, one could in principle expect insulating behavior to arise from an orbital ordering of the type $(d_{xy}d_{xz}d_{yz})^2 \rightarrow d_{xy}^0 d_{xz}^1 d_{yz}^1$.
However, while the compressive strain indeed splits the $t_{2g}$ levels, it also favors electron hopping in the \textit{xy} plane. This results in an increased bandwidth of the $d_{xy}$ orbital which counteracts the effect of the level splitting and does not allow for a gap to open, as shown in the bandstructure plotted in Fig.~\ref{fig:THEORY}(c).\\
\\
A reduction of the $d_{xy}$ bandwidth could be caused by a distortion of the in-plane Cr-O bonds through a collective tilting of the octahedral cages common to many perovskites. 
%
To test this, we initialize such octahedral tilts within our structures and then perform a subsequent relaxation for different values of the fixed in-plane lattice constant. This indeed results in an energy lowering compared to the undistorted $P4/mmm$ structure, and a stable distorted $Pbnm$ structure for the A-AF case, see solid orange line in Fig.~\ref{fig:THEORY}(a). For the C-AF case, the octahedral tilts are not stable and the system always relaxes back to $P4/mmm$.
\\
\\
We characterize the resulting distorted $Pbnm$ structures by two angles, $\phi$ and $\theta$, which measure the deviations of the out-of-plane Cr-O-Cr bond angle and of the in-plane O-O-O angle from 180$^\circ$ and 90$^\circ$, respectively (using the specific definition provided in Ref.~\cite{Dymkowski2014}).
These angles are plotted in Fig.~\ref{fig:THEORY}(b) for the A-AF phase as a function of the in-plane lattice constant.
Interestingly, we find that these octahedral tilts are stabilized by the Hubbard interaction $U$, in particular in the low compressive strain regime.
While for $U=0$ eV, the rotations are minor and the material remains metallic, for $U=2.5$ eV, the system sustains larger tilt angles ($\sim$ 3$^\circ$) and simultaneously exhibits an electronic band gap, as shown in Fig.~\ref{fig:THEORY}(d).
For very strong compressive strain, the out-of-plane tilt angle, $\phi$, and the bandgap are reduced, until they both eventually vanish around $a=3.61$ \AA.
We note that a stabilization of octahedral tilts by a Hubbard $U$ interaction has previously been reported also for another $d^2$ systems, namely the closely related compound SrMoO$_3$~\cite{Hampel2021}.\\
\\
We thus find that under compressive strain, an A-AF structure with small octahedral tilts becomes more favorable than the C-AF structure. Both the symmetry lowering and the reduced in-plane hopping due to the distorted Cr-O bonds, in combination with the local Hubbard interaction on the Cr ions, allow to open a bandgap between nominally occupied $d_{xz}$ and $d_{yz}$ and nominally empty $d_{xy}$ orbitals, resulting in an insulating band-structure, consistent with the resistivity measurement for the case of the \sco film grown on the \nao substrate ($1.82\%$ compressive strain).

\subsection*{Magnetism}
\begin{figure*}[tb]
    \centering
    \includegraphics[width=1\textwidth]{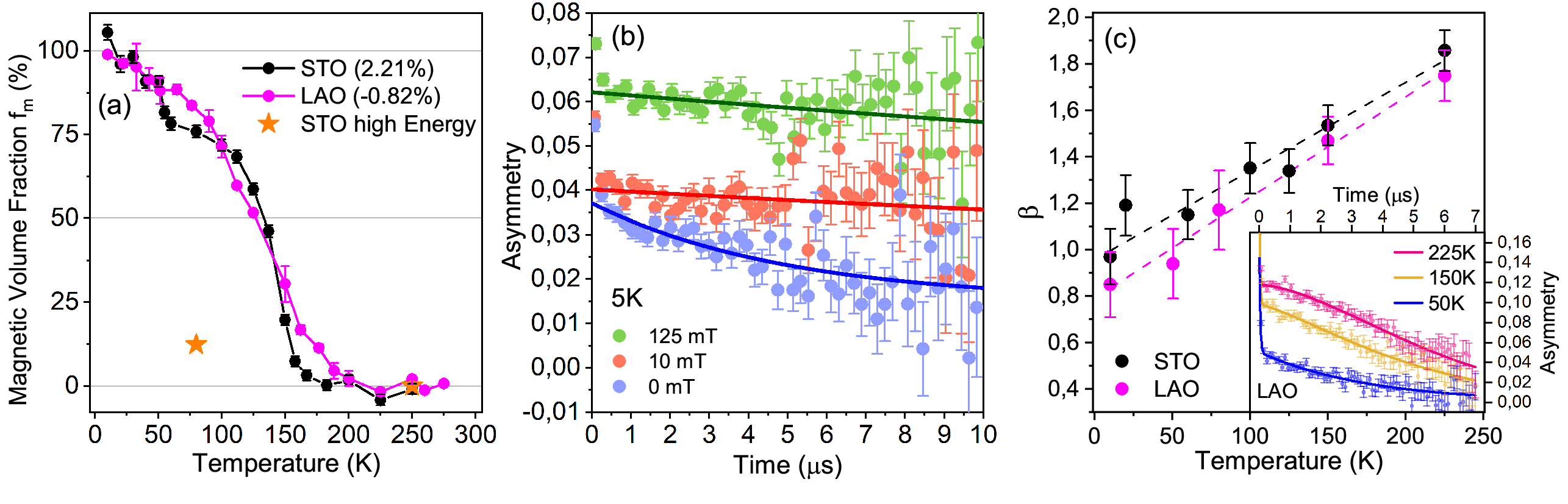}
    \caption{Magnetic muon spin relaxation (\msr) characterization of strained \sco thin films grown on \lao (pink) and \sto (black). 
    \textbf{(a)} Magnetic volume fraction $f_m$ extracted from $\mu$SR weak transverse field ($\SI{75}{\milli\tesla}$) measurements for the films grown on. The yellow star indicates the magnetic volume fraction for muons stopping predominantly in the substrate by using a higher implantation energy of $\SI{6.5}{\kilo\electronvolt}$ (see text). 
    \textbf{(b)} Muon asymmetry measurements in longitudinal field geometry, hence parallel to the muon spin direction, measured on the film grown on \lao.
    \textbf{(c)} $\beta$-exponent obtained by fitting zero field (ZF) \msr measurements with the formula \ref{eq:ZF_fit}. The inset shows the ZF asymmetry measured at $\SI{225}{\kelvin}$, $\SI{150}{\kelvin}$ and $\SI{50}{\kelvin}$. All ZF measurements are displayed in the SI Fig. \ref{figS:Muon_Supp}(c) and (d).
    }
    \label{fig:FIG3}
\end{figure*}
Having established the strain-dependent transport behaviour of \sco thin films, we turn now to muon spin relaxation (\msr) measurements to investigate their magnetic properties. 
In general, conventional magnetometry is inadequate for measuring antiferromagnetism in thin films because the contributions of the substrate and the film are difficult to identify and disentangle. 
\msr makes use of the asymmetric decay of the muon into a positron, which is emitted preferentially along the muon's spin direction.
This technique thus requires a spin-polarized muon beam, whose implantation depth depends on the muon's energy.
Immediately after implantation in the sample and before the said decay, however, the muon's spin has time to precess in the local magnetic field, which is then observed as a positron intensity oscillation. 
Consequently, the detection of positron intensity as a function of time enables probing the local magnetic field.
In the weak transverse field measurement mode (wTF), an applied magnetic field perpendicular to the initial muon spin direction induces a detectable spin precession in the non-magnetic phase, whereas the broad field distribution present in the magnetic phase leads to a fast depolarization.
Examples of the wTF asymmetry time curves for \sco films at different temperatures are shown in the SI Fig. \ref{figS:Muon_Supp}(b).
The extracted asymmetry oscillation amplitude $A(T)$ is then converted to magnetic volume fraction using 
\begin{equation}
    f_m(T) = (A_\text{max}-A(T))/(A_\text{max}-A_\text{bgr}) \cdot100\%
\end{equation}
where $A_\text{max}$ is the maximum amplitude in the non-magnetic regime and $A_\text{bgr}$ a known systematic background value. 
Figure \ref{fig:FIG3}(a) plots the temperature-dependent evolution of $f_m$ for two \sco films grown under compressive and tensile strain, on \lao and \sto, respectively, as obtained from $\SI{7.5}{\milli\tesla}$-wTF measurements.
At high temperatures, $f_m = 0$ is observed, which is expected for a pure paramagnetic phase. A steep increase in magnetic volume fraction occurs in both samples at around $\SI{150}{\kelvin}$, with $f_m$ reaching 100\% below $\SI{40}{\kelvin}$. Thus, \sco films undergo a magnetic transition at around $\SI{150}{\kelvin}$, independent of the epitaxial strain.
\\\\
Although a magnetic phase transition is clearly determined, the nature of the magnetic order cannot be inferred from these experiments. In principle, any static magnetic phase and spin fluctuations could also cause the increase in the magnetic volume fraction measured in wTF mode. 
Spin fluctuations can be ruled out by \msr using a longitudinal field (LF) applied parallel to the muon spin direction. In this configuration, the external field does not lead to a muon precession, and the asymmetry is expected to be constant over time, in the absence of magnetic fluctuations. LF-measurements performed at $\SI{5}{\kelvin}$ with $\SI{0}{\milli\tesla}$, $\SI{10}{\milli\tesla}$ and $\SI{125}{\milli\tesla}$ are presented in Figure \ref{fig:FIG3}(b). 
At $\SI{0}{\milli\tesla}$, a visible asymmetry decay is linked to the nuclear dipolar fields, which is discussed in detail in the ZF \msr analysis below.
This weak spin dynamics is nevertheless easily quenched by $\SI{10}{\milli\tesla}$ and $\SI{125}{\milli\tesla}$.
At these fields, an almost constant muon asymmetry is observed, implying that a purely static magnetic phase develops at low temperature, excluding the presence of spin fluctuations. Note that spurious spin dynamics remain detectable as a slight decrease in asymmetry.\\
\\
In-plane ferromagnetic order can be excluded, as no ferromagnetic signal was detected by SQUID magnetometry. Possible out-of-plane ferromagnetism was investigated using the wTF geometry with an implantation energy of $\SI{6.5}{\kilo\electronvolt}$. At this energy, the muon stopping depth is around $\SI{50}{\nano\meter}$, which is well below the film (see Figure \ref{figS:Muon_Supp}(a)).
This allows us to probe the absence of stray magnetic fields propagating from the film into the substrate, thus also excluding a ferromagnetic out-of-plane order. Since the stopping probability of the high-energy muons remains non-zero within the film, a finite magnetic volume fraction of 13\% is nonetheless still observed at $\SI{80}{\kelvin}$, as indicated by the yellow star in Figure \ref{fig:FIG3}(a). 
Additionally, net magnetic-moment-sensitive X-ray magnetic circular dichroism (XMCD) measurements were performed. 
Figure \ref{figS:XMCD}(a) shows the XMCD energy scans at the Cr L$_{2,3}$ edges of \sco films grown on \lao, \ngo, and \sto at $\SI{20}{\kelvin}$ as a function of the magnetic field. The strictly linear dependence of the XMCD signal on the applied magnetic field and the absence of saturation at high fields, as highlighted in Fig. \ref{figS:XMCD}(b), provide further evidence for the absence of ferromagnetic order in the \sco.
Finally, we provide further support for an antiferromagnetic spin order in the form of x-ray magnetic linear dichroism (XMLD) (see SI Fig. \ref{figS:XMLD}).\\
\\
All the measurements presented above therefore rule out the presence of a ferromagnetic spin order as well as substantial spin fluctuations as the origin of the magnetic phase transition observed by \msr,  
consistent with an AF-ordered ground state. Nevertheless, several observations motivate a more detailed investigation of the magnetic state. First, the magnetic transition $T_\text{Néel} = \SI{150}{\kelvin}$ appears unexpectedly high compared with prior measurements on bulk polycrystalline \sco, where $T_\text{Néel} = \SI{40}{\kelvin}$ \cite{ortega-san-martin_microstrain_2007}. Yet more intriguing is the very broad and seemingly strain-independent transition temperature. In perovskite systems, magnetic ordering, such as an AF ordering, is typically driven by superexchange interactions, which are highly sensitive to interatomic distances and bond angles. Visibly, the transport properties of the \sco films show a strong response to strain, and therefore, one could question the lack of such a response in the magnetic properties. 
Two possible causes are suggested here: a short-ranged dilute disordered AF state or a glassy state, both plausibly caused by remaining oxygen vacancies. 
A tool to discriminate between both scenarios is the analysis of the temperature-dependent ZF \msr data shown in the inset of Fig. \ref{fig:FIG3}(c) and in the SI Fig. \ref{figS:Muon_Supp}.
This data can be fitted with the stretched exponential function \cite{yaouanc_muon_2011, keren_probing_1996}
\begin{equation}
    \label{eq:ZF_fit}
    A(t) = A_Fe^{-\lambda_Ft} + A_Se^{(-\lambda_St)^{\beta}}
\end{equation}
where $A_F$ and $\lambda_F$ are the asymmetry and relaxation rate of the fast component (below $\SI{1}{\mu\second}$), while $A_S$ and $\lambda_S$ are for the slower-relaxing one. 
The temperature-dependent exponents $\beta(T)$ extracted from the fits, shown in Fig. \ref{fig:FIG3}(c), provide insight into the evolution of the magnetic environment. At high temperatures, $\beta \simeq 2$ indicates a Gaussian-like decay of the asymmetry. 
This case can be linked to the Kubo-Toyabe function, in which the local fields, arising from the dipolar fields of atomic nuclei, are Gaussian-distributed.
Upon cooling, $\beta$ decreases approaching $\beta \simeq 1$ near $\SI{0}{\kelvin}$. This exponential-like decay is indicative of disorder or slow spin fluctuations. Since $\beta$ remains above $0.8$, a disordered AF system is to be privileged over a canonical spin glass state, for which values  $0.3 \leq \beta \leq 0.5$ are typically reported \cite{hernandez-melian_muon-spin_2023, hatt_cluster_2025, campbell_dynamics_1994, keren_probing_1996}. However, $\beta$ alone cannot conclusively distinguish between these scenarios, and additional field- and temperature-dependent \msr measurements in the wTF mode would be required. 
\\\\
Assuming a dilute disordered AF system in our \sco thin films, the percolative formation of magnetically ordered regions, instead of a clear magnetic transition, is a plausible scenario. 
As the temperature decreases, the AF ordered regions gradually grow, resulting in a very broad transition.
In fact, prior studies of perovskite systems, such as $\ce{La2CuO4}$ or $\ce{SrMnO3}$, have shown that epitaxial strain causes the Néel transition to shift by only $\SI{10}{\kelvin}$ to $\SI{40}{\kelvin}$ by comparing samples grown on \lao and \sto. \cite{suter_antiferromagnetic_2004, maurel_nature_2015}.
If \sco exhibits a similar strain-induced shift in transition temperature, it may be difficult to distinguish from our \msr data.
Despite these limitations, our measurements, as well as the DFT calculations, support the presence of an AF spin order with a dilute magnetic disorder in \sco films under a wide range of epitaxial strains.

\section{Conclusion}
In conclusion, we have successfully grown high-quality \sco thin films on a variety of substrates, imposing epitaxial strains ranging from $-1.82$ to $3.4\%$ and thereby considerably expanding the strain range explored for this material. 
Transport measurements confirm the metallic character of the ground state of \sco, exemplified by films grown on \lao and \ngo, and reveal a strain-induced metal-to-insulator transition at approximately $1.2\%$. Whereas the insulating behavior in the tensile-strain regime is associated with the emergence of orbital order accompanied by a Jahn-Teller distortion, further DFT calculations indicate that rotations and tilts of the CrO$_6$ octahedra can likewise stabilize an insulating state under compressive strain. 
Using a combination of $\mu$SR and XMCD/XMLD measurements, we demonstrate the presence of AF order at low temperatures with an onset of the magnetic transition at $\SI{150}{\kelvin}$. The nearly strain-independent magnetic transition is attributed to dilute magnetic disorder, which promotes the percolative formation of AF-ordered regions and broadens the transition. 
Overall, these findings establish the coexistence of metallic conductivity and AF order, confirming that the ground state of \sco thin films is a metallic antiferromagnet. 
More broadly, our results provide a comprehensive framework for understanding the interplay between lattice, orbital, electronic, and magnetic degrees of freedom in \sco, and motivate further investigations of perovskite alkaline-earth chromates to elucidate the microscopic mechanisms governing their correlated electronic and magnetic properties.

\section{Description of Methods}
\subsection*{Experimental methods}
All \sco thin films were grown by radio-frequency off-axis magnetron sputtering with a sputtering power of $\SI{35}{\watt}$. The substrates were kept at $\SI{750}{\celsius}$ during growth in an atmosphere composed of a mixture of oxygen and argon with an O$_2$-to-Ar ratio of $0.01\%$. This atmosphere was held at a pressure of $\SI{0.045}{\milli\bar}$ under constant flow. The oxygen partial pressure was intentionally kept low to avoid over-oxidising the film during the growth. 
To ensure high crystalline quality, $5$ to $6$ unit cells of $\ce{SrTiO_3}$ were grown as a buffer layer using the same growth conditions. No structural relaxation or influence on the physical properties has been observed due to the buffer layer. After growth, the films are annealed at $\SI{250}{\celsius}$ for 5 hours with a constant $\SI{2}{\liter\per\hour}$ oxygen flow. The necessity of this procedure is described by Zhang et al. \cite{zhang_reversible_2014}. 
All substrates were provided by CrysTec GmbH, Germany, and to ensure a high-quality surface with step-like morphology, the substrates were annealed between $\SI{1000}{\celsius}$ and $\SI{1100}{\celsius}$ with an oxygen flow before growth.\\
\\
X-ray diffraction (XRD) measurements were performed using a Rigaku SmartLab diffractometer using a monochromated X-ray source ($\SI{1.5406}{\angstrom}$, Cu K-$\alpha_1$ edge). 
The lattice parameter and film thickness were determined from the XRD patterns by fitting them using the InteractiveXRDFit MATLAB tool \cite{lichtensteiger_interactivexrdfit_2018}. 
Atomic force microscopy (AFM) images were performed in non-contact mode using a Park Systems NX10 device and evaluated using the GWYDDION software. Temperature-dependent transport measurements were performed using a Quantum Design Physical Properties Measurement System (PPMS) and the thin film samples were patterned to a Hall-bar geometry by photolithography. 
The scanning transmission electron microscopy images were obtained with a double-corrected Thermofisher Spectra 300 scanning transmission electron microscope, operated at $\SI{200}{\kilo\volt}$, using a high angular annular dark field detector and a convergence semiangle of 20mrad. A thin lamella of the sample was prepared using a Thermofisher Helios-UX focused ion beam (FIB). The measurements and preparations were performed at ICN2.
The magnetic properties were investigated primarily by muon spin relaxation (\msr) conducted at the LEM endstation of the S$\mu$S facility at the Paul Scherrer Institut (PSI), Villigen, Switzerland. 
To maximize the \msr signal, six identical $5\times\SI{5}{\square\milli\meter}$ \sco film samples were tiled together in a mosaic. If not stated otherwise, a low kinetic energy ($\SI{1.5}{\kilo\electronvolt}$) corresponding to a mean penetration depth of about $\SI{10}{\nano\meter}$ was used. 
All \msr data have been analyzed by musrfit \cite{suter_musrfit_2012} 
X-ray magnetic circular/linear dichroism (XMCD/XMLD) measurements have been conducted at the X-Treme endstation at the Swiss Light Source facility at PSI, Switzerland \cite{piamonteze_x-treme_2012}, and at the BOREAS endstation at the ALBA synchrotron facility, Spain \cite{barla_design_2016}.
These measurements have been performed in the total electron yield (TEY) mode with grazing incidence ($\SI{60}{\degree}$ to normal) and magnetic fields up to $\SI{6}{\tesla}$.
The XMCD signal is defined as the difference in absorption of left and right-circular polarised X-ray light. The XMCD signal magnitude in Fig. \ref{figS:XMCD} is obtained from the XMCD intensity difference between the on-resonant (L$_3$-edge) and off-resonant (background) energy at each applied magnetic field.
\subsection*{Computational methods}
DFT calculations are performed using the Quantum Espresso package v7.1~\cite{Giannozzi2009_QE_general}.
We use the generalized gradient approximation in the form by Perdew, Burke, and Ernzerhof for solids (PBEsol)~\cite{Perdew2009_PBEsol} as exchange-correlation functional together with the GBRV ultrasoft pseudopotential library~\cite{Garrity2014_GBRV_pseudos}.
All calculations are performed for a ($\sqrt{2} \times \sqrt{2} \times 2$) supercell of the primitive perovskite unit cell, which can accommodate both A-AFM and C-AFM.
We further use a ($10 \times 10 \times 7$) \textit{k}-point grid, employing the Marzari-Vanderbilt type ``cold-smearing'' \cite{Marzari1999_smearing}, a kinetic energy cutoff of 52~Ry for the plane wave coefficients, and a charge density cutoff of 624~Ry.\\
\\
To better account for the Coulomb repulsion in the Cr $d$-states, we employ the Hubbard $U$ correction (DFT+$U$)~\cite{Anisimov1991_original_DFTU, Cococcioni2005_QE_DFTU} where we use orthonormalized atomic-like (\textit{i.e.} ortho-atomic) projectors to apply the $+U$ correction. We underline that the use of ortho-atomic projectors instead of non orthonormalized atomic ones is the main reason why we observe some quantitative but not qualitative differences with respect to the already published results~\cite{carta_evidence_2022}.\\
\\
We perform epitaxially constrained cell relaxations by fixing the in-plane lattice constants while allowing the out-of-plane lattice constant and all internal atomic coordinates to relax freely. This procedure is applied to three structural types: an undistorted tetragonal structure with $P4/mmm$ symmetry; a distorted structure exhibiting octahedral tilts and rotations ($a^-a^-c^+$ in Glazer notation) with $Pbnm$ symmetry; and a Jahn-Teller distorted structure characterized by $P2_1/a$ symmetry and C-AFM magnetic order~\cite{carta_evidence_2022}. While this $P2_1/a$ cell is capable of accommodating both Jahn-Teller modes and octahedral tilts, we note that the tilts are not favorable and vanish during the relaxation process.
%
%
%
%
%
\section*{Acknowledgement}
The authors acknowledge Izabela Bia\l{}o for fruitful discussions, and thank Jean-Marc Triscone and Lucia Varbaro for their help and for kindly providing access to their laboratory facilities for patterning the samples. This research was supported by the Swiss National Science Foundation
(SNSF) under Project No. PP00P2$\_$170564. S.J. and M.G. acknowledge support from the Agility plus grant from MARVEL NCCR (SNSF Project No. 51NF40182892). 
G.D.L. acknowledges grant RYC2021-032524-I funded by MCIN/AEI/10.13039/501100011033 and by “European Union NextGenerationEU”/PRTR.
Many thanks to A.P. for her contribution during her semester project. 
A.C. and C.E. were supported by the NCCR MARVEL, funded by the Swiss National Science Foundation (Grant No. 182892)
DFT calculations were performed on the Euler cluster of ETH Zurich.
We thank J. H-M. for his support at the BOREAS beamline (proposal no. 2023027342) and C.P. for her support at the X-treme beamline.
For the STEM images, we thank for the support of B.M. and Marcos Rosado (proposal no. 20240320053). The authors acknowledge the use of instrumentation as well as the technical advice provided by the Joint Electron Microscopy Center at ALBA (JEMCA) and funding from Grant IU16-014206 (METCAM-FIB) to ICN2 funded by the European Union through the European Regional Development Fund (ERDF), with the support of the Ministry of Research and Universities, Generalitat de Catalunya. The authors acknowledge TU Wien Bibliothek for financial support through its Open Access Funding Programme. 
\section*{Authors Contributions}
S.J. and M.G. conceived the project and organized the experiments. The samples were grown and characterized by S.J. with the help of J.S., A.P. and M.G. 
XAS, XMLD and XMCD measurements were performed by S.J., J.S., G.D.L and M.G. with the support of C.P. at the SLS, PSI, and J.H, B.M. at ALBA.
All $\mu$SR measurements were conducted and analyzed by A.S.
DFT calculations were performed by A.C. under the supervision of C.E.
The manuscript was written by S.J. and  M.G. with collaborative contributions from all authors.
\section*{Conflicts of interest}
The authors declare no conflict of interest.

\bibliography{Misc/Lib_SJ_AC_2}
\bibliographystyle{Misc/MSP.bst}
\end{multicols}

\section{Supplementary Information}
\renewcommand{\thefigure}{S\arabic{figure}}
\setcounter{figure}{0}
\begin{figure}[!h]
    \centering
    \includegraphics[width = 0.8\textwidth]{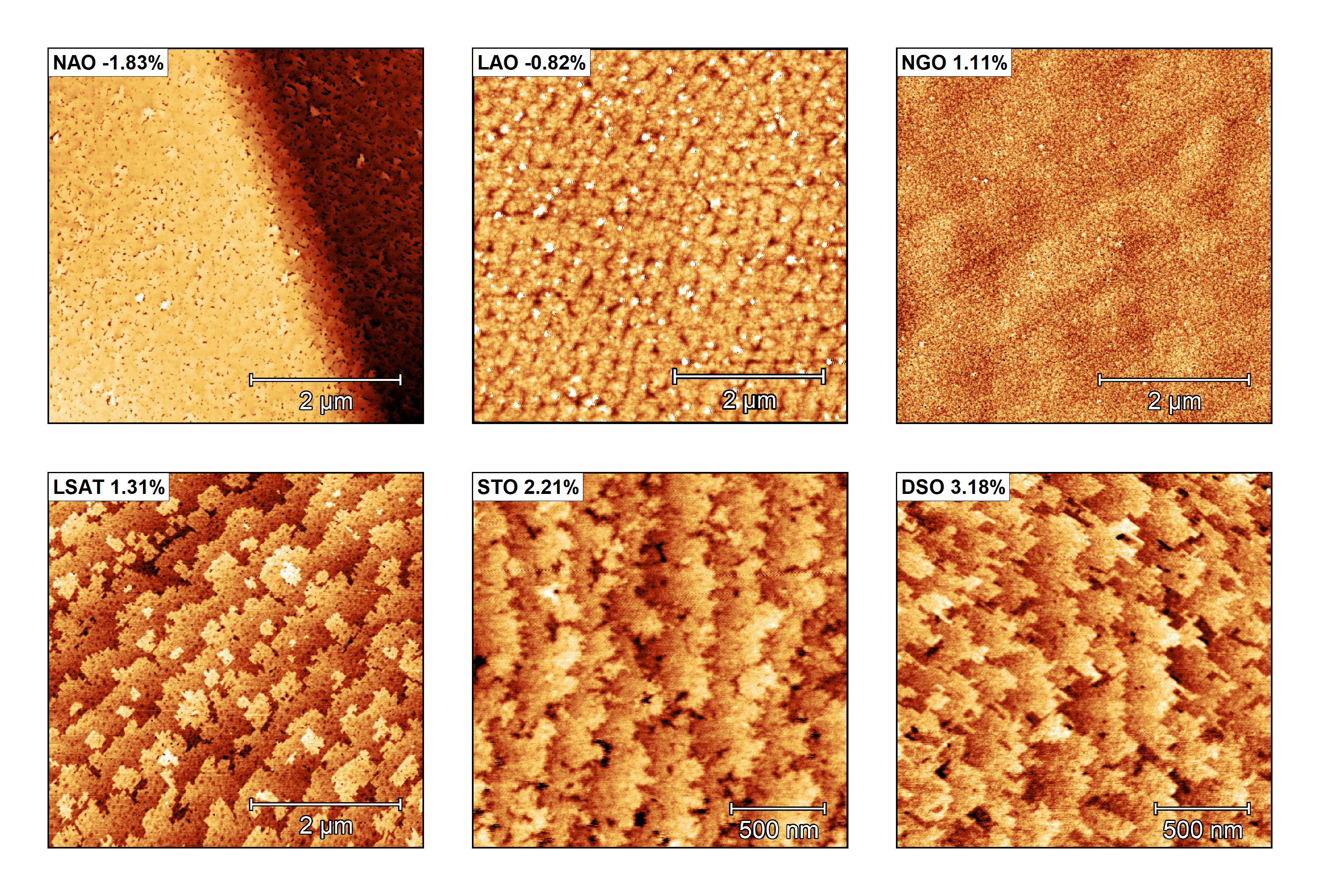}
    \caption{Atomic force microscopy images of strained \sco thin films grown on various substrates. The substrate type and the associated strain values are indicated. The scanned area is $5\times5\mu$\SI{}{\square\meter}}
    \label{figS:AFM}
\end{figure}
%
%
\begin{figure}[!h]
    \centering
    \includegraphics[width = 0.8\textwidth]{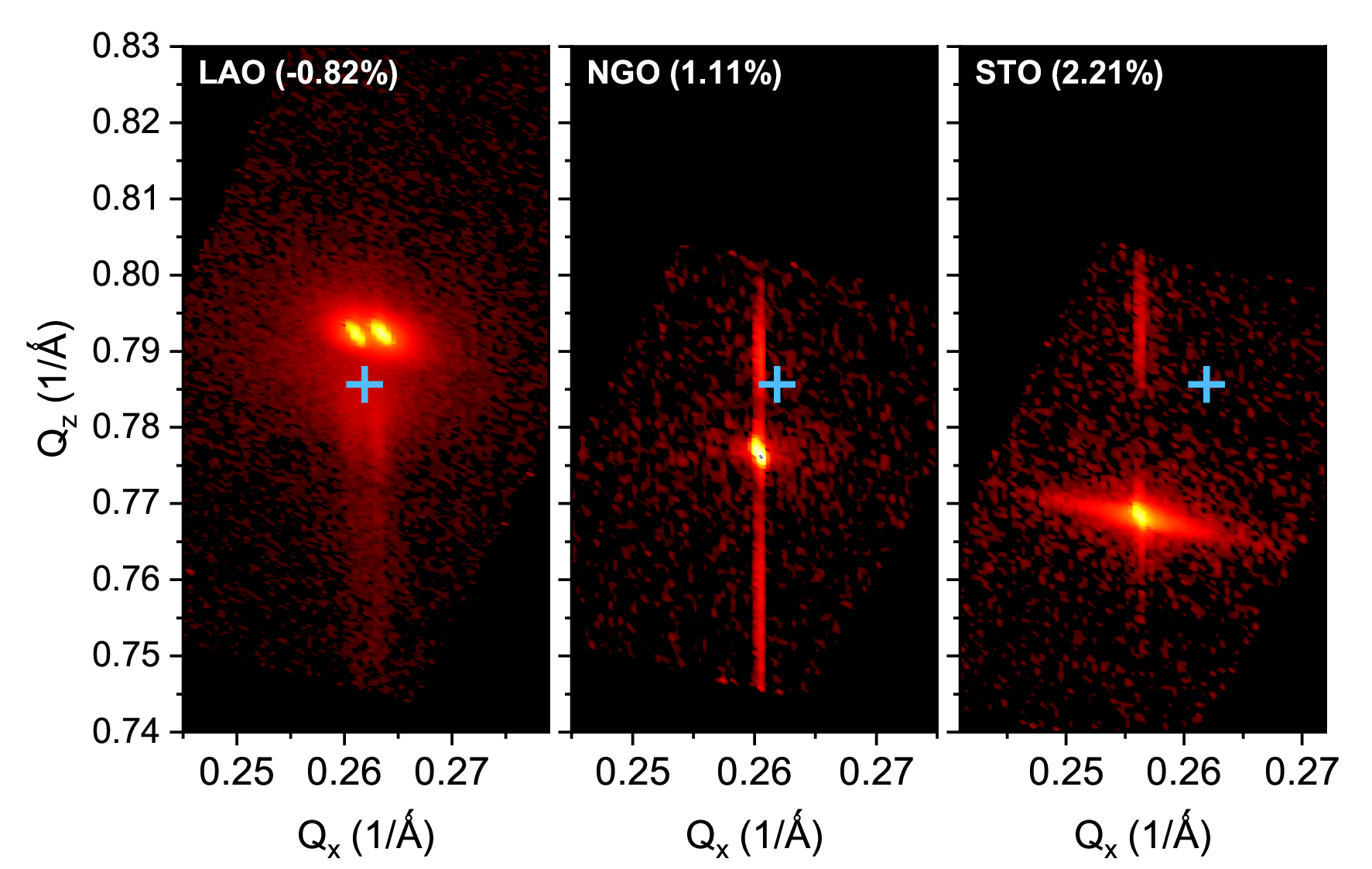}
    \caption{Reciprocal space maps measured on three strained \sco films. The blue cross represents the position of bulk \sco.}
    \label{figS:RSM}
\end{figure}
\begin{figure}[h]
    \centering
    \includegraphics[width = 0.6\textwidth]{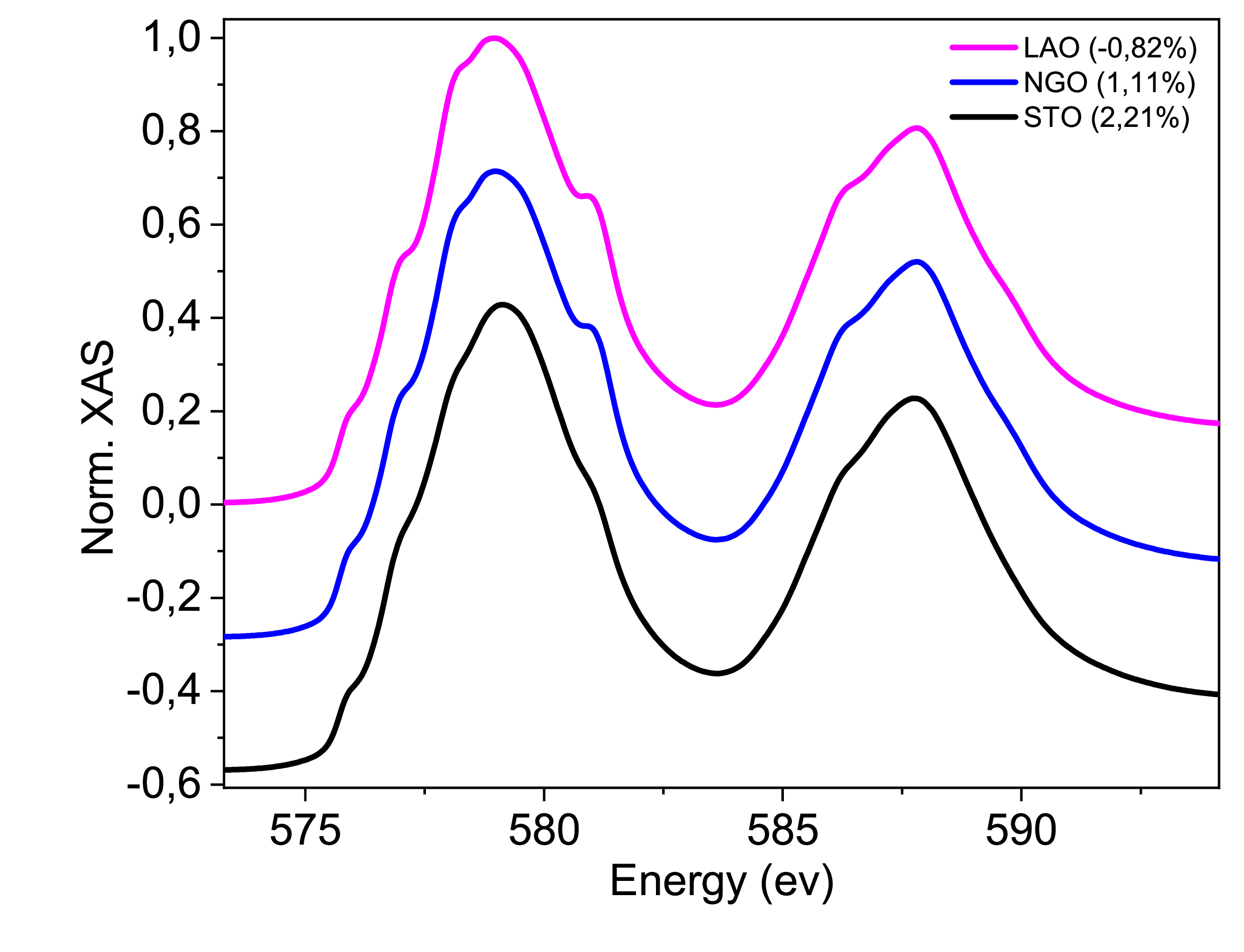}
    \caption{Room temperature XAS performed on \sco grown on \lao, \ngo and \sto}
    \label{figS:XAS}
\end{figure}
%
\begin{figure}[h]
    \begin{center}
    \includegraphics[width = 0.9\textwidth]{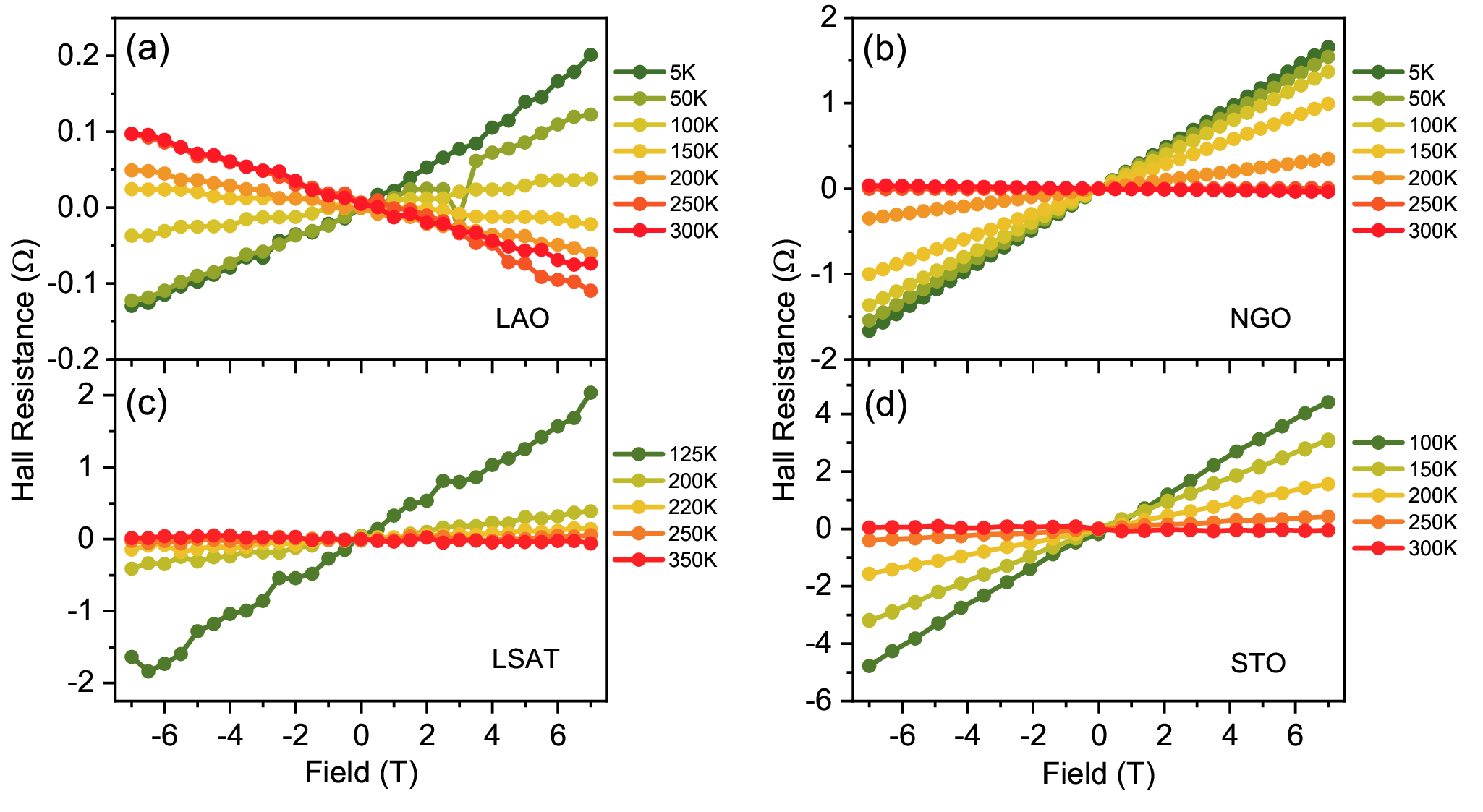}
    \caption{Hall resistance as a function of the applied magnetic field and temperature for \sco films grown on (a) \lao, (b) \ngo, (c) LSAT and (d) \sto.}
    \label{figS:Hall}
    \end{center}
The data presented in Figure \ref{figS:Hall} does allow us to calculate the hall coefficients (Fig. \ref{fig:FIG3}(b)) and consequently the charge carrier density at the Fermi edge. So at $\SI{5}{\kelvin}$ we find $n = \SI{1.729e22}{\per\cubic\centi\meter}$ and $\SI{2.275e21}{\per\cubic\centi\meter}$ for the \lao and \ngo samples, respectively. 
In the LSAT and \sto samples, because of the high resistivity, only the carrier density at $\SI{100}{\kelvin}$ could be estimated, where we find $n=\SI{1.170e21}{\per\cubic\centi\meter}$ and $\SI{1.016e21}{\per\cubic\centi\meter}$, respectively. 
Compared to typical values of $n$ for metals \cite{ashcroft_solid_1976}, the \lao and \ngo samples are both in a similar range, down to very low temperatures, confirming their metallic nature. These values were used to calculate the critical resistivity using the Ioffe-Regel relationship. \\
\\
In fact, the Ioffe-Regel criterion assumes that a system is strongly localized when the inverse Fermi-momentum $k_F$ is smaller than the ballistic mean free path $l$ of the charge carrier, in other words, $k_F l<1$. Using the Drude model for transport and this Ioffe-Regel criterion, one can determine the critical resistivity 
\begin{equation}
    \rho_c = \frac{\hbar k_F^2}{e^2 n}
\end{equation}
where $n$ is the charge carrier density and the Fermi-momentum $k_F$, which can be expressed using the relation $\frac{8\pi}{3} \frac{1}{(2\pi)^3} k_F^3 = n$.
This equation, together with the carrier density values determined from the Hall effect measurements, allows us to calculate the resistivity limit for each strained film.
\end{figure}

%
\begin{figure}[h]
    \centering
    \includegraphics[width = 0.8\textwidth]{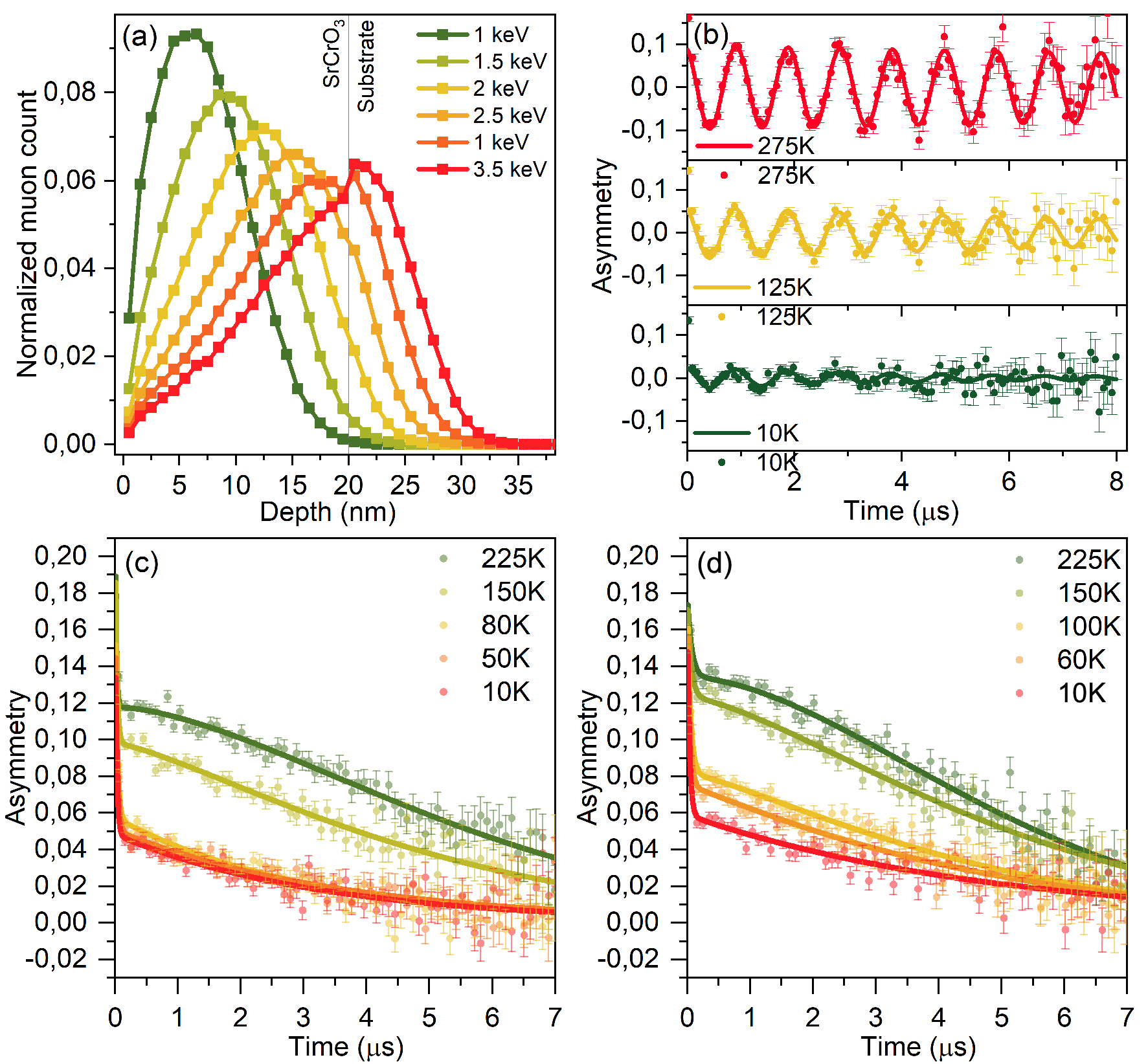}
    \caption{\textbf{(a)} Simulated normalized muon stopping depth distribution as a function of the muon energy \cite{eckstein_computer_1991}. The simulation is made assuming a $\SI{20}{\nano\meter}$ thick \sco layer with a density of $\SI{5.62}{\gram\per\cubic\centi\meter}$ on a \lao  substrate of $\SI{6,51}{\gram\per\cubic\centi\meter}$. \textbf{(b)} Weak transverse field  (wTF) \msr measurements showing the muon asymmetry at selected temperatures, above and below the transition $T_\text{Néel} = \SI{150}{\kelvin}$. \textbf{(c)} and \textbf{(d)} Zero field \msr asymmetry measurements as a function of the temperature for films grown on \lao (c) and \sto (d). The transition from a Gaussian-like decay curve at high temperature to an exponential decay of the asymmetry at low temperature clearly shows the magnetic transition leading to a broad distribution of local magnetic fields.}
    \label{figS:Muon_Supp}
\end{figure}
\begin{figure}[h]
    \centering
    \includegraphics[width=0.9\linewidth]{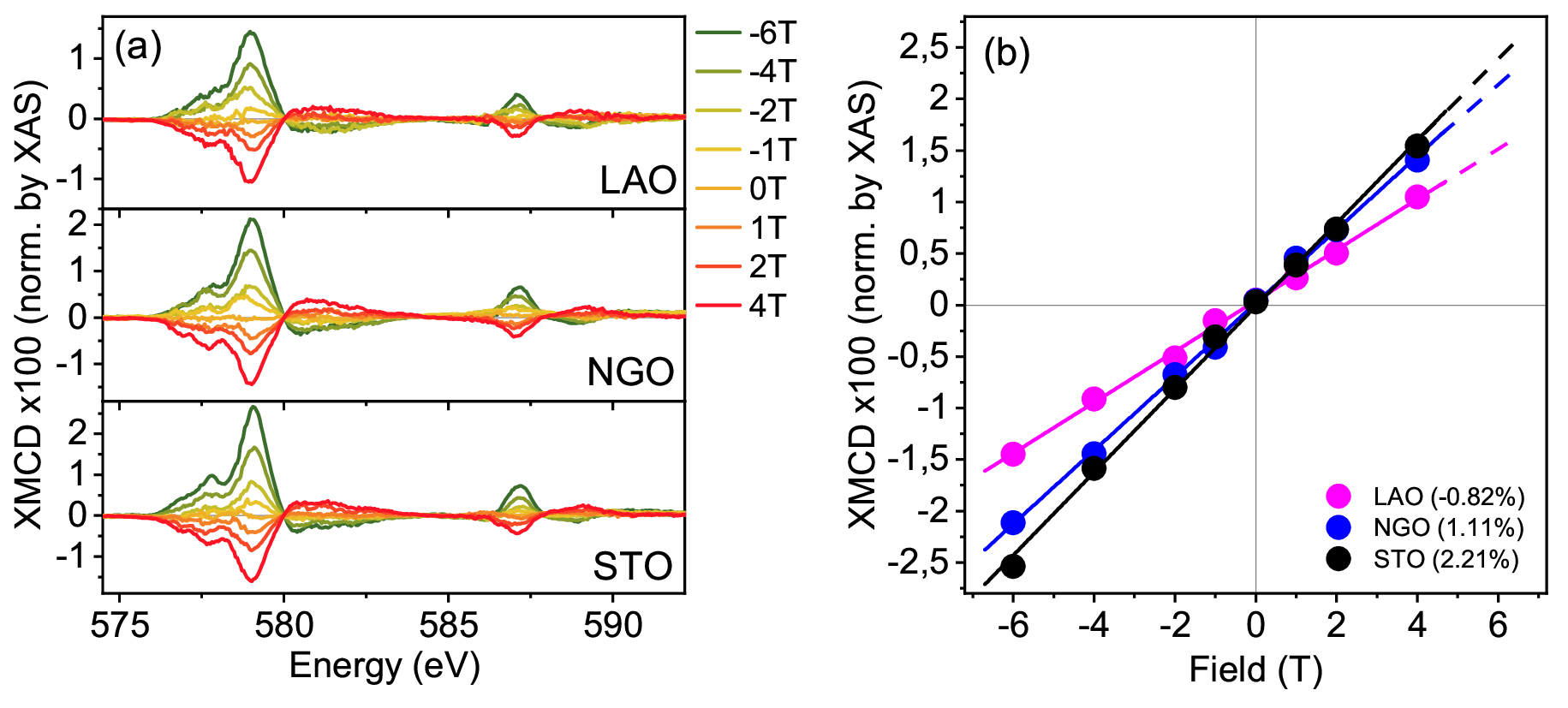}
    \caption{XMCD measurements performed at $\SI{20}{\kelvin}$ on \sco thin films grown on \lao, \ngo and \sto as a function of the field. \textbf{(a)} XMCD energy scans of the $\ce{Cr}$-L$_2$ and L$_3$ edges. \textbf{(b)} XMCD intensity as a function of field. The dashed lines are linear guides to the eye. For technical reasons, we were unable to measure XMCD for the $+\SI{6}{\tesla}$ case.}
    \label{figS:XMCD}
\end{figure}
\begin{figure}[h]
    \begin{center}
    \includegraphics[width = 1\textwidth]{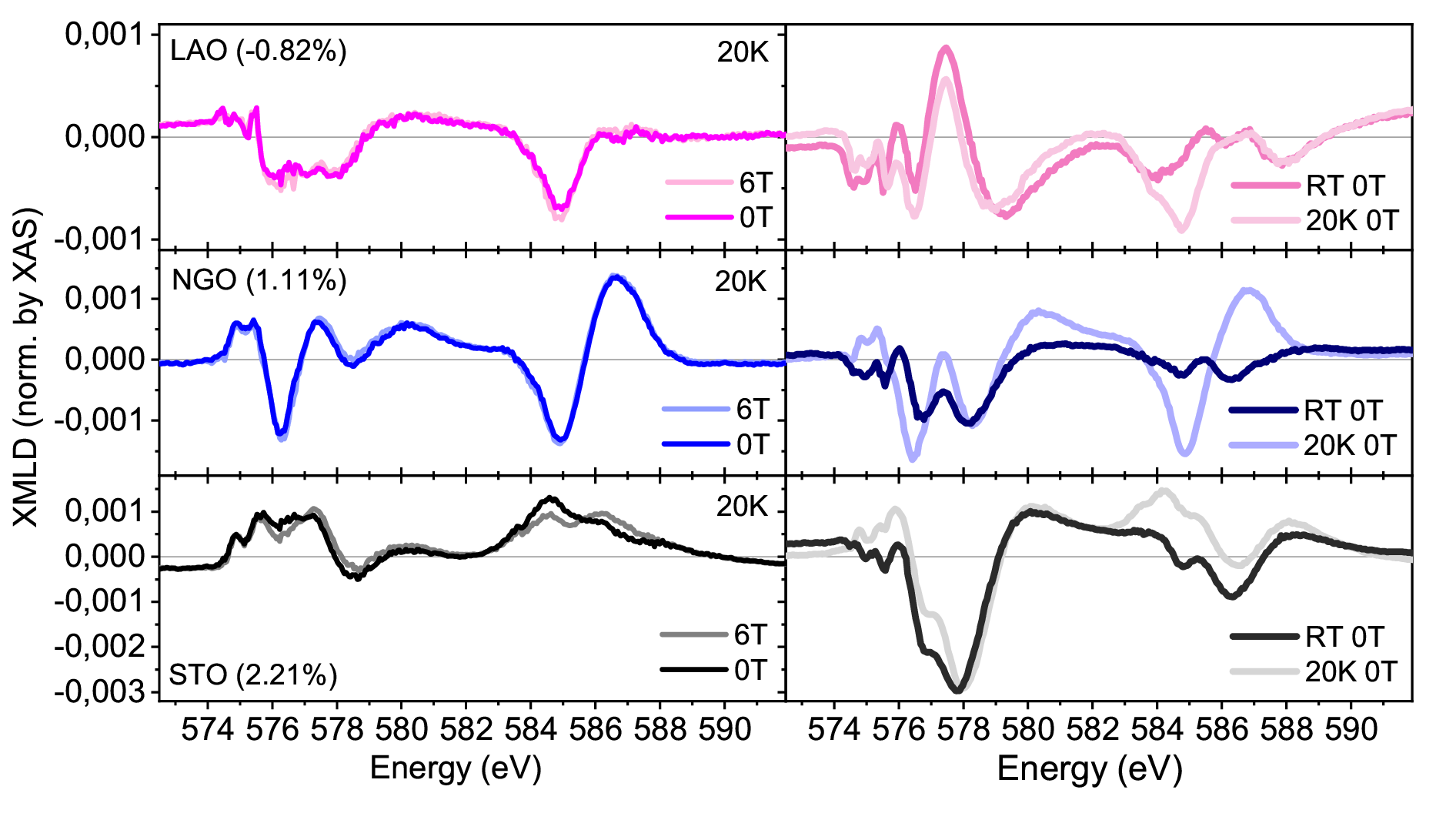}
    \caption{\textbf{Left colon} X-ray magnetic linear dichroism (XMLD) of \sco films on \lao, \ngo and \sto measured at $\SI{20}{\kelvin}$ without and with $\SI{6}{\tesla}$ of external magnetic field at grazing incidence. \textbf{Right colon} Uncorrected XMLD signal measured at room temperature (RT) and $\SI{20}{\kelvin}$ without magnetic field.}
    \label{figS:XMLD}
    \end{center}
Support for an antiferromagnetic spin order is derived from x-ray magnetic linear dichroism (XMLD) (See Fig. \ref{figS:XMLD}).
Assuming that the orbital contribution of the XMLD signal does not change significantly with temperature, the magnetic contribution at low temperature can be obtained by subtracting the XMLD signal measured at room temperature, i.e., XMLD$_{\SI{20}{\kelvin}}-$XMLD$_{\SI{300}{\kelvin}}$  \cite{aruta_orbital_2009}. The XMLD is sensitive to both a ferromagnetic and an antiferromagnetic signal, but they can be disentangled by applying a high enough external magnetic field \cite{haverkort_magnetic_2004, gibert_interfacial_2015}. Our data appears to be consistent with an antiferromagnetic order as there are only negligible differences between $\SI{0}{\tesla}$ and $\SI{6}{\tesla}$ scans.
\end{figure}

\end{document}